\documentstyle[epsfig,psfig]{mn}     
     
\newif\ifAMStwofonts     
     
 \input epsf.tex    
     
\ifoldfss     
  \ifCUPmtlplainloaded \else     
    \NewTextAlphabet{textbfit} {cmbxti10} {}     
    \NewTextAlphabet{textbfss} {cmssbx10} {}     
    \NewMathAlphabet{mathbfit} {cmbxti10} {} 
    \NewMathAlphabet{mathbfss} {cmssbx10} {} 
  \fi     
  \ifAMStwofonts     
    \ifCUPmtlplainloaded \else     
      \NewSymbolFont{upmath} {eurm10}     
      \NewSymbolFont{AMSa} {msam10}     
      \NewMathSymbol{\upi}     {0}{upmath}{19}     
      \NewMathSymbol{\umu}     {0}{upmath}{16}     
      \NewMathSymbol{\upartial}{0}{upmath}{40}     
      \NewMathSymbol{\leqslant}{3}{AMSa}{36}     
      \NewMathSymbol{\geqslant}{3}{AMSa}{3E}

      \let\leq=\leqslant      
            
    \fi     
  \fi     
\fi 
     
\ifnfssone     
  \newmathalphabet{\mathit}     
  \addtoversion{normal}{\mathit}{cmr}{m}{it}     
  \addtoversion{bold}{\mathit}{cmr}{bx}{it}     
  \newmathalphabet{\mathbfit} 
  \addtoversion{normal}{\mathbfit}{cmr}{bx}{it}     
  \addtoversion{bold}{\mathbfit}{cmr}{bx}{it}     
  \newmathalphabet{\mathbfss} 
  \addtoversion{normal}{\mathbfss}{cmss}{bx}{n}     
  \addtoversion{bold}{\mathbfss}{cmss}{bx}{n}     
  \ifAMStwofonts     
    \ifCUPmtlplainloaded \else     
      \UseAMStwoboldmath     
      \makeatletter     
      \new@mathgroup\upmath@group     
      \define@mathgroup\mv@normal\upmath@group{eur}{m}{n}     
      \define@mathgroup\mv@bold\upmath@group{eur}{b}{n}     
      \edef\UPM{\hexnumber\upmath@group}     
      \new@mathgroup\amsa@group     
      \define@mathgroup\mv@normal\amsa@group{msa}{m}{n}     
      \define@mathgroup\mv@bold\amsa@group{msa}{m}{n}     
      \edef\AMSa{\hexnumber\amsa@group}     
      \makeatother     
      \mathchardef\upi="0\UPM19     
      \mathchardef\umu="0\UPM16     
      \mathchardef\upartial="0\UPM40     
      \mathchardef\leqslant="3\AMSa36     
      \mathchardef\geqslant="3\AMSa3E     

      \let\leq=\leqslant      

    \fi     
  \fi     
\fi 
     
\ifnfsstwo     
  \DeclareMathAlphabet{\mathbfit}{OT1}{cmr}{bx}{it}     
  \SetMathAlphabet\mathbfit{bold}{OT1}{cmr}{bx}{it}     
  \DeclareMathAlphabet{\mathbfss}{OT1}{cmss}{bx}{n}     
  \SetMathAlphabet\mathbfss{bold}{OT1}{cmss}{bx}{n}     
  \ifAMStwofonts     
    \ifCUPmtlplainloaded \else     
      \DeclareSymbolFont{UPM}{U}{eur}{m}{n}     
      \SetSymbolFont{UPM}{bold}{U}{eur}{b}{n}     
      \DeclareSymbolFont{AMSa}{U}{msa}{m}{n}     
      \DeclareMathSymbol{\upi}{0}{UPM}{"19}     
      \DeclareMathSymbol{\umu}{0}{UPM}{"16}     
      \DeclareMathSymbol{\upartial}{0}{UPM}{"40}     
      \DeclareMathSymbol{\leqslant}{3}{AMSa}{"36}     
      \DeclareMathSymbol{\geqslant}{3}{AMSa}{"3E}     

      \let\leq=\leqslant      

    \fi     
  \fi     
\fi 
     
\ifCUPmtlplainloaded \else     
  \ifAMStwofonts \else 
    \def\upi{\pi}     
    \def\umu{\mu}     
    \def\upartial{\partial}     
  \fi     
\fi

\title{The synchrotron foreground and CMB temperature--polarization cross
correlation power spectrum from the first year WMAP data}
    
\author[G. Bernardi et al.]     
       {G. Bernardi,$^{1}$ E.~Carretti,$^1$ R.~Fabbri,$^2$ C.~Sbarra,$^1$     
        S.~Cortiglioni$^1$\\     
        $^1$C.N.R./I.A.S.F. Bologna, Via Gobetti 101,      
         I-40129 Bologna, Italy\\     
        $^2$Dipartimento di Fisica, Universit\`a di Firenze, Via Sansone 1,     
            I-50019 Sesto Fiorentino (FI), Italy\\     
}     
\date{24 October 2002}     
     
\pagerange{\pageref{firstpage}--\pageref{lastpage}}     
\pubyear{0000}

\begin{document}

\maketitle

\label{firstpage}

\begin{abstract} 
We analyse the temperature-polarization cross-correlation in the Galactic synchrotron template that we recently developed, and between the template and  CMB  temperature maps derived 
from WMAP data. Since the polarized
synchrotron template itself uses WMAP data, we can 
estimate residual synchrotron contamination in the CMB   
$C_{\ell }^{TE}$  angular spectrum. While $C_{2 }^{TE}$
appears to be contaminated by synchrotron, no evidence for 
contamination is found in the multipole range which is most relevant for
the fit of the cosmological  optical depth.   
\end{abstract}     


\section{Introduction}

\label{intro} The WMAP experiment measuring the temperature-polarization
cross-correlation power spectrum of the Cosmic Microwave Background (CMB)
found an excess of power at large angular scales ($\ell <10$), which has
been interpreted as evidence for an early reionization (Kogut et al. 2003). 
A clean measurement of the cosmological signal relies on a successful
removal of the foregrounds, which on large angular scales are mainly generated 
by dust, free-free and synchrotron emissions from the Galaxy. In particular,
the synchrotron radiation is the main polarized foreground at WMAP
frequencies. According to Bennett et al. (2003) the CMB maps used to compute
the angular power spectrum $C_{\ell }^{T}$ have negligible foreground
contamination, thanks to the wide frequency coverage of the WMAP experiment
and a safe foreground subtraction achieved with fits of foreground templates.
Also Kogut et al. (2003) claimed that the contamination in the Q,
V and W bands is low when the Galactic plane is cut out and the $C_{\ell
}^{TE}$ power spectrum of the CMB is free of foreground contamination.

However, several groups have performed independent analyses of the WMAP data
to address the foreground contamination on the CMB maps. Tegmark, de
Oliveira-Costa \& Hamilton (2003, hereafter TOH) claimed to have obtained
a CMB map\footnote{%
http://www.hep.upenn.edu/~max/} cleaner than the one of the WMAP team. Naselsky et
al. (2003) applied a phase analysis to the internal linear combination map
obtained by the WMAP team, showing some residual foreground contamination.
Also, Naselsky et al. (2004) compared the analysis of the internal
linear combination map obtained by the WMAP team's analysis with TOH's, and
found evidence for a residual contamination in the low--multipole power
spectrum region. Dineen \& Coles (2004a, 2004b) used the cross--correlation
between the rotation measures of extragalactic radio sources and the CMB
maps to identify a possible foreground residual; they found evidence for
that in both the WMAP and TOH CMB maps. However, these works cannot tell us 
whether the foreground residual may affect the $C_{\ell }^{TE}$ power
spectrum to a significant extent, and it looks harder to improve the
cross-correlation analysis of Kogut et al. (2003) because polarization maps
have not been provided yet by WMAP. The issue of possible foreground
contamination on $C_{\ell }^{TE}$ \ is however very important in the light
of the reported anomalies in WMAP's large-scale output, including
North-South asymmetries (Eriksen et al. 2004, Hansen et al. 2004b, Land and
Maguejo 2005) and multipole alignments (TOH, de Oliveira-Costa
 et al. 2004, Copi et al. 2003). As far as we are concerned about the
robustness of \ the $C_{\ell }^{TE}$ power spectrum and the inferences on
cosmological reionization, the most troublesome result is due to Hansen et
al. (2004a), according to which the high optical depth ascribed to the
cosmological medium by WMAP's team should originate from the Southern
(Galactic or Ecliptic) emisphere.

In the light of the above results, it is necessary to investigate further
the possible impact of foregrounds on the $C_{\ell }^{TE}$ \ power spectrum.
\ Due to the absence of measured polarization maps at microwave frequencies,
uncertainties on the foreground contamination in polarization are greater
than in total intensity. In this $Letter$ we want to tackle the problem of
the foreground contamination on the CMB $C_{\ell }^{TE}$ power spectrum on
large angular scales using the templates of synchrotron polarized emission
developed by Bernardi et al. (2003) and Bernardi et al. (2004, hereafter
B04). The B04 template in particular is expected to be much more accurate
than the earlier one which relies on the extrapolation of surveys at $\sim $
1 GHz. \ However, its superiority derives from the use of WMAP's total
intensity maps. 

\section{Synchrotron and CMB $C_{\ell }^{TE}$ power spectra}

\begin{figure}
\includegraphics[width = 1\hsize]{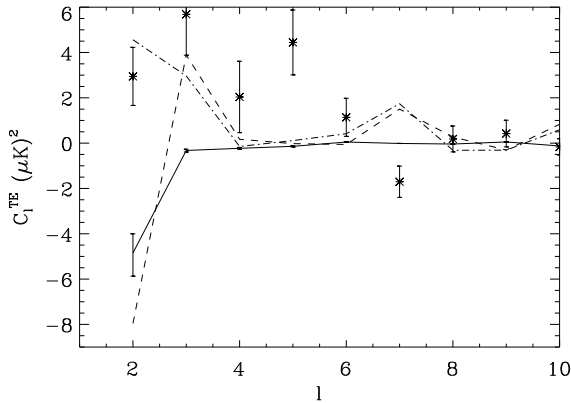}
\caption{ The synchrotron $C_{\ell }^{TE}$ power spectrum at 60~GHz
(solid line) compared to the CMB $C_{\ell }^{TE}$ 
from Kogut et al. (2003) (asterisks). Also reported are
the $C_{\ell }^{T_{CMB}E_{SYNCH}}$ power spectra for the WMAP
CMB map (dashed line) and
for the TOH CMB map (dash--dotted).}
\label{te_spectrum}
\end{figure}

B04 provided $Q$ and $U$ template 
maps of the Galactic synchrotron emission at
23~GHz, which is obtained from the 23~GHz total
intensity synchrotron map released by the WMAP team 
(by using the polarization direction field of starlight as well as 
a polarization horizon model). B04 also derived a Galactic map of the synchrotron spectral index,
which is used to scale the polarization template to higher frequencies. In
the present work, the same spectral index map is used to scale the WMAP
23-GHz total intensity synchrotron map released by the WMAP team. 
We choose a frequency of  60~GHz, which can be regarded as an
approximate mean of the frequencies of the WMAP QVW data set.
Synchrotron $T$, $Q$ and $U$  maps (with $T$ the antenna temperature) 
are generated at an angular
resolution of $7^{\circ }$ due to the limitation of the B04 template.
We are then able to compute the 60-GHz synchrotron 
$C_{\ell }^{TE}$ power spectrum
by integration of the two-point
correlation function. This procedure (see, for the implementation, Sbarra et
al. 2003) allows to properly account for the incomplete sky coverage of the
B04 template, and for the kp2 Galactic-Plane mask applied to the synchrotron
maps (for details, see Bennett et al. 2003). A similar procedure allows us
to investigate 
cross-correlations between the CMB and the synchrotron template. 

Figure~\ref{te_spectrum} shows the synchrotron $C_{\ell }^{TE}$ power
spectrum at 60~GHz and the CMB $C_{\ell }^{TE}$ power spectrum measured by
WMAP. The error bars on the synchrotron spectrum only account for a
variation $\Delta \alpha =\pm 0.2$ of the frequency spectral index of the
synchrotron emission. Therefore they represent the uncertainty on the
overall normalization of the fiducial synchrotron template at 60~GHz, and do
not account for statistical errors. Clearly, the 60-GHz synchrotron $C_{\ell
}^{TE}$ is much smaller than the corresponding CMB spectrum in the range $%
\ell =3-10$ where most information about cosmological reionization is
encoded. The situation of course\ would be still better at 90 GHz. We note
that the synchrotron quadrupole $C_{2}^{TE}$ has a large and negative value
and this indicates a potential source of contamination for the CMB
quadrupole; however, this very fact does not indicate any inadequacy in
WMAP team's technique of foreground removal. On the other hand, 
reasonably strong
evidence for a residual contamination can be provided by a cross-correlation
between the CMB temperature and the synchrotron polarization fields, both
being derived from WMAP's data.

In order to cross--correlate our $Q$ and $U$ \ templates with CMB
anisotropy, we use two different CMB maps. The first one is obtained by
averaging the Q, V and W maps released by the WMAP team after foreground
subtraction (we refer to this as the WMAP CMB map); the other one is the CMB
map produced by TOH. The $C_{\ell }^{T_{CMB}E_{SYNCH}}$ power spectra
computed for both these maps are also shown in Fig.~\ref{te_spectrum}. \
Both power spectra show very similar behaviours for $\ell >2.$ We find no
evidence of a CMB-synchrotron correlation in the range $\ell =4-10$. The
multipole $\ell =3$ shows a cross-correlation which at 60 GHz is comparable
to the CMB $C_{3}^{TE}$. This 
should not be so disturbing after all, since the
large reionization optical depth is essentially generated by 
slightly larger $\ell$'s. 
The most intriguing feature is still the behaviour of the quadrupole.
When the synchrotron template is correlated with the WMAP CMB map we find a
large ($\sim 8$~$\mu $K$^{2}$) negative value, whereas the use of the TOH
CMB map leads to a relatively lower ($\sim 5$~$\mu $K$^{2}$) but positive
value. For comparison, the CMB quadrupole is $C_{2}^{TE}\sim 3$~$\mu $K$^{2}$%
, a factor 2--4 lower than the magnitude of $C_{2}^{T_{CMB}E_{SYNCH}}$.

\begin{figure}[tbp]
\centerline{\epsfxsize=9cm
\epsfbox%
{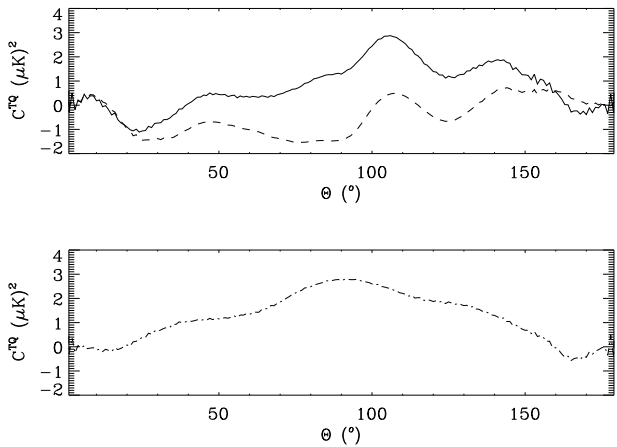}}
\caption{Top panel: The correlation function between the synchrotron $Q^{r}$ 
and the CMB temperature maps provided by WMAP's team  (solid line) 
and by TOH (dashed line). Bottom panel: The difference between the two
correlation functions.}
\label{corr_funz}
\end{figure}

The discrepancy between cross-correlation quadrupoles is not surprising,
since TOH already noted that their temperature quadrupole $C_{2}^{T}$ is
significantly different from the one found by the WMAP team. This
discrepancy, as well as the overall behaviour of the cross-correlation power
spectra, is better understood by inspection of the CMB-synchrotron
cross-correlation function. The latter is defined by 
\begin{equation}
C^{TQ}(\theta )=\sum_{ij}I_{i}Q_{j}^{r},
\end{equation}%
where  the Stokes parameter  $Q^{r}$ is computed 
in the frame of the great circle
connecting pixels $(i,j)$. Figure~\ref{corr_funz} shows the
cross-correlation functions derived from WMAP and TOH maps and their
difference. 
It is interesting to note that the cross-correlation functions have the same
behaviour up to $\theta \sim 20^{\circ }$ but they differ significantly for
larger angular scales. There is a clear evidence for strongly
correlated signals between the WMAP CMB map and the polarized synchrotron at
angular scales $\theta >50^{\circ }.$ The absolute maximum of $C^{TQ}(\theta
)$ is $\sim 2$~$\mu $K$^{2}$ for $\theta \sim 110^{\circ }$. The
cross-correlation function derived from the TOH CMB map is somewhat weaker,
and it has an opposite sign at several scales, especially the scales that
are the most relevant in the determination of $C_{2}^{T_{CMB}E_{SYNCH}}$.
However, the difference between the two correlation functions approximately
has a very simple angle dependence $\Delta C^{TQ}(\theta )$ $\propto \sin
\theta $ and is as large as $\sim 2\mu $K$^{2}$~for $\theta \sim 90^{\circ }$%
. This fully accounts for the discrepancy found for $%
C_{2}^{T_{CMB}E_{SYNCH}} $. \ The rather moderate decrease of both $%
C^{TQ}(\theta )$ and \ $C_{2}^{T_{CMB}E_{SYNCH}}$ achieved by using the TOH
temperature map seems to imply that a residual contamination by synchrotron
emission survives on the largest angular scales even in the TOH treatment.

\section{Conclusions}

The main purpose of this \textit{Letter} is the study of the possible
synchrotron contamination of the CMB $C_{\ell }^{TE}$ power spectrum derived
from the first year WMAP release. The comparison between the CMB and
the synchrotron template
$C_{\ell }^{TE}$ power spectra shows that the contamination in
the 60-GHz CMB $C_{\ell }^{TE}$ should be negligible for $3\leq \ell \leq
10. $ The inspection of the cross-correlation spectrum $C_{\ell
}^{T_{CMB}E_{SYNCH}}$ (as well as the analysis of the cross-correlation
function) reinforces this conclusion at least in the range $4\leq \ell \leq
10$. On the other hand, the high values of the synchrotron quadrupole $%
C_{2}^{TE}$ and of the  CMB-synchrotron cross-correlated signal on large angular
scales ($\theta >50^{\circ })$ suggest a residual synchrotron
contamination. We emphasize that the B04 template uses the WMAP total
intensity synchrotron map at \ 23 GHz: The CMB contamination on that map is
not likely to correlate so strongly with the
best CMB maps available today, after rescaling to 60 GHz. 
Therefore, the contamination should be in the
CMB maps; this would also explain why the synchrotron template is more
correlated with the WMAP CMB map than with the TOH one. The relatively high
level of contamination might be, potentially, a serious problem for CMB
quadrupole $C_{2}^{T}$. Although this point is far from being proved by the
present analysis, the foreground contamination might partly account
for the low CMB quadrupole measured by COBE and WMAP.

In spite of the above open problem, we find that the multipole region where
there is no evidence for contamination  includes the range which
dominates the standard WMAP fitting of the reionization optical depth. This
result, obtained at 60 GHz, should be adequately representative for the QVW
data set. Therefore, it seems really hard to explain in this way the
North-South asymmetry in the optical depth fits declared by Hansen et al.
(2004a).  If such an asimmetry is confirmed, it should be of extragalactic 
(although not necessarily cosmological) origin.  This possibility is
supported by Schwartz et al. (2004)
in connection with other WMAP anomalies.

\smallskip \textit{Acknowledgments:} This work has been carried out in the
frame of the SPOrt programme funded by the Italian Space Agency (ASI).

\bsp     

\label{lastpage}

\end{document}